\documentclass{elsart4}

\usepackage{amssymb}

\usepackage{graphics}
\usepackage{graphicx}
\usepackage{epsfig}
\usepackage{amssymb}
\journal{Physica B}

\begin{document}

\begin{frontmatter}

\title{Similarities in electronic properties of organic charge-transfer solids and layered cobaltates \thanksref{label1}}
\thanks[label1]{Supported by DOE Grant No. DE-FG02-06ER46315.}
\author{Sumit Mazumdar,\thanksref{add1}\corauthref{cor1}}
 \author{~R.~Torsten~Clay,\thanksref{add2}}
 \author{Hongtao.~Li \thanksref{add1}}
 \address[add1]{Department of Physics, University of Arizona, Tucson, AZ 85721, USA}
\address[add2]{Department of Physics and Astronomy and HPC$^{2}$ Center for Computational Sciences, Mississippi State University, Mississippi State, MS 39762, USA}
 \corauth[cor1]{Tel: +001-520-621-6803; FAX:+001-520-621-4721; e-mail: sumit@physics.arizona.edu}





\begin{abstract}
The apparently counterintuitive carrier concentration-dependent
electronic properties of layered cobaltates have attracted wide
interest. Here we point out that very similar carrier-concentration
dependence has previously been noted in strongly correlated quasi-one
dimensional (quasi-1D) organic charge-transfer solids. The normal
states of both families can be understood, over the entire range of
carrier concentration of interest, within the extended Hubbard
Hamiltonian with significant intersite Coulomb interaction. As with
the charge-transfer solids, superconductivity in the cobaltates
appears to be limited to bandfilling of one-quarter. We point out
further that there exist other families of correlated superconductors,
such as spinels, where too strong correlations, geometric lattice
frustration and bandfilling of one-quarter seem to be the essential
features of the unconventional superconductors.
\end{abstract}

\end{frontmatter}

\section{Introduction}
Twenty five years after the discovery of high T$_c$ superconductivity
(SC) in the cuprates, none of the proposed scenarios have led to a
consistent mechanism of this phenomenon. 
It is now
recognized that there exist many other superconductors in which
electron-electron (e-e) interactions are repulsive. One approach to
arriving at the theory of correlated-electron SC 
is to determine
the common characteristics shared by different families of
correlated-electron superconductors. Once such common features are
determined, one could ask what the implications of these features are
for the mechanism of SC. It is with this goal we report and explain
the strong parallels between conducting organic charge-transfer solids
(CTS) and layered cobaltates. Both families have been compared
individually to the cuprates.  Resonating Valence Bond (RVB) theories
of SC have in the past been proposed for both quasi-2D
$\kappa$-(BEDT-TTF)$_2$X \cite{Powell06a} and the layered hydrated
cobaltate Na$_x$CoO$_2$ $\cdot$ $y$H$_2$O, $x \sim 0.35$
\cite{Baskaran03a,Kumar03a,Motrunich04a}.

Our description here uses the carrier concentration $\rho$, rather
than the bandfilling. In the CTS, $\rho$ is the charge per individual
molecule.  
The charge carriers in the CTS occupy the highest molecular orbitals of
the molecules, which act as single sites. The layered cobaltates, -
Na$_x$CoO$_2$, Li$_x$CoO$_2$ and K$_x$CoO$_2$ - consist of CoO$_2$
layers separated by layers of alkali ions. The Co ions form a
triangular lattice and have average charge (4-$x$)+. The large crystal
field splitting \cite{Wang03a} leads to low-spin states for the
Co-ions, and the Co$^{3+}$ (Co$^{4+}$) ions are spinless (spin
$\frac{1}{2}$.)  Trigonal distortion splits the $t_{2g}$ $d$-orbitals
on the Co ions further into two low-lying $e^{\prime}_g$ orbitals and
a higher $a_{1g}$ orbital, and photoemission studies of Na$_x$CoO$_2$
suggest that the charge carrying holes on the Co$^{4+}$ ions occupy
the $a_{1g}$ orbitals only \cite{ARPES,Laverock07a}. {\it Note that
  the hole density $\rho=1-x$ here.}

In both families $\rho$ can be tuned over wide ranges. We briefly
describe the experimental observations, and then present a consistent
theory of the {\it systematic} $\rho$-dependence in the two apparently
unrelated families. Following this we point out that SC in the CTS and
the cobaltates may be occurring at the same $\rho=0.5$.  Given the
strong role of carrier concentration 
in both families, this 
cannot be a coincidence. There exist other inorganic materials where also SC is
limited to the same $\rho$. 
A correct theory of
correlated-electron SC 
should explain 
this shared feature.

\section{Experimental Observations}

In the quasi-1D CTS $\rho$ ranges from 0.5 to 1. The $\rho=1$
materials are Mott-Hubbard semiconductors \cite{Torrance75a}.
Experimental signatures of strong correlations in the $\rho < 1$
conductors are, (i) magnetic susceptibility $\chi(T)$ enhanced
relative to the calculated Pauli susceptibility $\chi_P$, and (ii)
4k$_F$ instability. {\it Structurally similar 1D conductors, with
  nearly identical molecular components and crystal structures,
  exhibit very different behavior.} Thus $\chi(300K)/\chi_P$ is $\sim$
20 in MEM(TCNQ)$_2$, $\sim3$ in TTF-TCNQ, and $\sim1$ in HMTSF-TCNQ
\cite{Mazumdar86a}.  The 4k$_F$ transition temperature $T_{4k_F}=335$K
in MEM(TCNQ)$_2$ \cite{Huizinga79a}, but considerably lower in
TTF-TCNQ \cite{Comes79a}. HMTSF-TCNQ shows no 4k$_F$ instability and
only the ``normal'' 2k$_F$ instability \cite{Comes79a}.  These 
observations had led to various ``large U'' and ``small U''
Hubbard model-based theories 
with, however, no understanding why U could be both large and small. 
More recently, the
emphasis has been on the 4k$_F$ charge-ordering in 2:1 cationic CTS \cite{Nad06a}.


$\chi(T)$ in Na$_x$CoO$_2$ is strongly $x$-dependent. Early work
\cite{Foo04a} labeled $\chi(T)$ as ``Pauli paramagnetic'' for $x<0.5$
(large $\rho$) and ``Curie-Weiss'' for $x>0.5$ (small $\rho$).  More
recent works label the small $x$ (large $x$) materials as ``weakly''
(``strongly'') correlated, and put the boundary between
these close to $x=0.67$ instead of 0.5 \cite{RecentCoO2a}.
Thermopower measurements give similar results. Early
theories ascribed the $x$-dependence of the electronic behavior in
Na$_x$CoO$_2$ to Na-ion potentials. Experiments \cite{RecentCoO2a}
that have established identical $x$-dependence in Li$_x$CoO$_2$ with
much smaller Li-ions and in the incommensurate misfit cobaltates show
that the behavior is intrinsic to the CoO$_2$ layers.  Weakly
correlated behavior for systems with $\rho$ close to 1, - which for
large enough $U$ would be a Mott-Hubbard semiconductor, - and strongly
correlated behavior for small $\rho$, where the system is closest to
being a band insulator with all Co-ions as spinless Co$^{3+}$,-
are both counterintuitive. Weak
correlation has sometimes been ascribed to greater mixing of $a_{1g}$
and $e_g^\prime$ orbitals, but it is not clear why such mixing should
be $x$-dependent.

\section{Theory}

The problems in understanding the $\rho$-dependence arise from
limiting discussions of e-e interactions to the simple Hubbard model,
with the onsite repulsion $U$ as the only significant Coulomb
interaction. The simple Hubbard model provides a good description of
the ground state and the low energy spin excitations for $\rho=1$,
where the effects of longer range Coulomb interactions can be
incorporated with an effective onsite interaction $U_{\rm{eff}}$ (with only
short range nearest neighbor (NN) repulsion $V$, $U_{\rm{eff}}\simeq U-V$.)
This effective parametrization breaks down for $\rho \neq 1$. We
demonstrate that the observed $\rho$-dependence in both CTS and
cobaltates can be understood very well within the extended Hubbard
Hamiltonian 
%
%
provided $U$ is {\it finite} and $V/U$ is nonnegligible.

We consider the Hamiltonian,
\begin{equation}
H=-\sum_{\langle ij \rangle \sigma}t_{ij}c^\dagger_{i\sigma}
c_{j\sigma} + U\sum_{i} n_{i,\uparrow}n_{i,\downarrow} 
+ V \sum_{\langle ij \rangle} n_{i} n_{j},
\label{ham1}
\end{equation}
for both the CTS and the cobaltates.  In the above $\langle ij
\rangle$ imply NN, the Fermion operator $c^\dagger_{i\sigma}$ creates
an electron or a hole with spin $\sigma$ ($\uparrow$ or $\downarrow$)
on a CTS molecular site, and a hole in the triangular lattice of
Co-ions in the layered cobaltates. All other terms have their usual
meanings.  In the following we express all energies in units of $|t|$.
As discussed in Section 2, the susceptibility relative to the 
Pauli susceptibility is the most used measure of the degree of
correlation. 
Since calculations of thermodynamic quantities can be done only with relatively small number of
electrons, we choose to calculate the normalized probability of double occupancy in the ground state,
\begin{equation}
g(\rho)=\frac{\langle n_{i,\uparrow}n_{i,\downarrow}\rangle}{\langle n_{i,\uparrow}\rangle \langle n_{i,\downarrow}\rangle}.
\end{equation}
which can be calculated for much larger system size. Explicit calculations of 
$g$ \cite{Mazumdar86a} and susceptibility \cite{Mazumdar83a} indicate that small (large) $g$ is both necessary and
sufficient for
enhanced (unenhanced) susceptibility.
The physical reason for this is that
the reduction of Hubbard Hamiltonians to the  Heisenberg spin Hamiltonian 
requires $g\rightarrow0$, while large $g$ implies metallic behavior.

We have performed exact numerical calculations of $g(\rho)$.  Our
previous 1D calculations \cite{Mazumdar86a,Mazumdar83a} were with different system
sizes for different $\rho$. Substantive improvements in computer
capabilities in the intervening years now allow us to vary $\rho$
keeping the same number of sites $N$. Our calculations are for
periodic rings of $N=16$.  In Fig.~1 we have plotted $g(\rho)$ against
$\rho$ for $U=10$, $V=0$, 2 and 3. The value of $U$ chosen is
realistic for CTS \cite{Torrance75a}. Although $\rho < 0.5$ systems do
not exist here we have retained one such a point for comparison to the
2D calculations below.
The key points of Fig.~1 are: (i) for $V=0$ correlation
effects are nearly independent of $\rho$, (ii) for $V \neq 0$,
correlations are strongest for materials near $\rho=0.5$, and (iii)
materials with $\rho$ between $\sim$ 0.67 - 0.9 should exhibit weakest
correlations.
\begin{figure}
\resizebox{2.8 in}{!}{\includegraphics{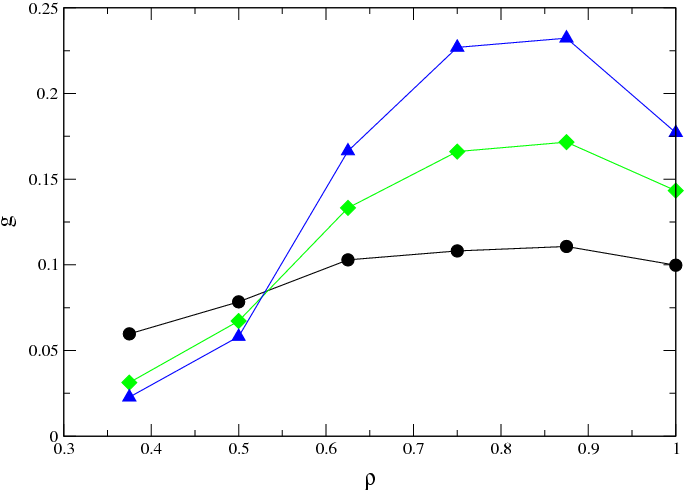}}
\caption{(Color online) Exact $g(\rho)$ versus $\rho$ for $U=10$, and
  $V=0$ (circles), 2 (diamonds) and 3 (triangles), for periodic ring
  of 16 sites. Lines are guides to the eye.}

\label{(Fig1)}
\end{figure}


Our calculations in 2D are for six different finite periodic
triangular lattices, with $N$=12, 16 18 and 20 (two different 16-site
and 20-site lattices each can be constructed) \cite{Li11a}. Here we
report results for the $N$=16 and 20-site lattices shown in Fig.~2(a)
and (b), respectively.  The corresponding plots for $g(\rho)$, for the
same $U$ and $V$ as in Fig.~1, are shown in Figs.~2(c) and (d),
respectively.  
These $U$ and $V$ were arrived at from experiments \cite{Li11a,Choy07a}.  
Importantly, $\rho$-dependent $g(\rho)$ in either 1D or 2D
is not found in the $U\rightarrow\infty$ limit, where $g \to 0$ for
all $\rho$. Finite $U$ and
$V$ are both required \cite{Mazumdar83a,Li11a}.
The natures of the plots for the two
different lattice sizes are very similar: (a) $g(\rho)$ is nearly
$\rho$-independent for $V=0$, (b) is small for $\rho \leq \frac{1}{3}$
for $V\neq0$, and (c) fairly large in the region $\rho > 0.4$ for
significant $V$. In particular, $V$ decreases $g(\rho)$ in the region
$\rho \leq \frac{1}{3}$ but enhances it for larger $\rho$. 
$\rho=\frac{1}{3}$ in the triangular lattice thus corresponds to $\rho=0.5$
in 1D. 
\begin{figure}
\resizebox{3.0 in}{!}{\includegraphics{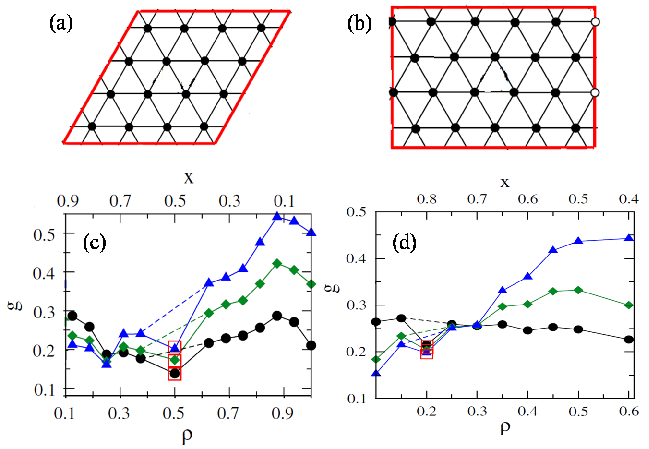}}
\caption{(Color online) (a) and (b) $N=16$ and 20-site clusters
  investigated numerically.  (c) and (d) Exact $g(\rho)$ versus
  $\rho$. The circles, diamonds and triangles correspond to the same
  parameters as in Fig.~1. The boxes correspond to data points with
  total spin $>S_{min} = 0(\frac{1}{2})$ for even (odd) numbers of
  particles.}
\label{g(rho2d)}
\end{figure}

\section{Comparison with experiments}

Fig.~2 predicts that 1D systems with $\rho=0.5$ and realistic $V$ are
the most strongly correlated, and the effective correlations decrease
with increasing $\rho$, until about $\rho \sim 0.8$ beginning from
where the effective correlations increase again. Precisely such a
systematic $\rho$-dependent behavior is seen in the entire CTS family
\cite{Mazumdar83a}.  Among the three CTS mentioned above, (i)
MEM(TCNQ)$_2$ exhibits very strongly correlated behavior because it is
$\rho=0.5$, where $g(\rho)$ is the smallest; (ii) TTF-TCNQ with
intermediate $\rho=0.59$ exhibits moderately strongly correlated
behavior, with $g(\rho)$ still smaller than that at $\rho=1$; (iii)
the weakly correlated behavior of HMTSF-TCNQ with $\rho=0.75$ is
expected, since $g(\rho)$ here close to being largest (see Fig.~1).
Recent discovery of 4k$_F$ charge ordering in the (TMTTF)$_2$X \cite{Nad06a}
and the accompanying theoretical discussions \cite{Monceau01a} are in agreement with 
these conclusions. 
Thus weakly and strongly correlated behavior can both emerge in spite
of having the same $U$ and $V$.

Figs. 2(c) and (d) predict that correlation effects in the triangular
lattice are strongest for $\rho \leq 0.33$ ($x>0.67$), and that the
boundary between strongly and moderately correlated regions occurs at
$\rho \sim 0.3-0.4$ 
Asymmetry between
$\rho=\frac{1}{3}$ and $\frac{2}{3}$ is predicted, with
$\rho=\frac{2}{3}$ predicted to show weakly correlated behavior.
In the $U \to \infty$ limit $\sqrt{3}\times \sqrt{3}$ charge-ordering
is expected in both cases \cite{Motrunich04a}.
Our predictions are in strong agreement with observations in
Na$_x$CoO$_2$, Li$_x$CoO$_2$ and the misfit cobaltates
\cite{RecentCoO2a}.  Here we have ignored the $e_g^\prime$
orbitals. 
Elsewhere \cite{Li11a} we have shown that 
the two-band extended (but not simple) Hubbard Hamiltonian explains the
weak $e_g^\prime$ hole occupancy near $\rho=0.67$, also in agreement with
experiments \cite{Laverock07a}.

\section{Superconductivity}

SC in correlated-electron superconductors invariably occurs over a
narrow range of carrier concentration, and sometimes the
superconducting substance is a line compound.  The latter is true with
superconducting CTS, which are 2:1 cationic compounds
($\rho=\frac{1}{2}$ hole carriers) or 1:2 anionic compounds
($\rho=\frac{1}{2}$ electron carriers).  We find this significant,
given the strong $\rho$-dependence of the electronic behavior 
and the natural explanation of the same within the
extended Hubbard model. Beyond this, superconducting CTS are quasi-2D
with strong interstack interactions.  Highest T$_c$ is reached in the
$\kappa$-(BEDT-TTF)$_2$X, in which the superconducting state is often
proximate to an antiferromagnetic state. The $\kappa$-lattice is
strongly dimerized, and the dimer unit cells form an anisotropic
triangular lattice. Theoretical explanation of the antiferromagnetism
requires that the dimer lattice be thought of as {\it effective}
$\rho=1$, with each dimer as a single site. This is what has led to to
the RVB theories of SC within the $\rho=1$ Hubbard Hamiltonian on an
anisotropic triangular lattice \cite{Powell06a}. Within these models
increasing pressure increases frustration as well as bandwidth, and SC
appears at the interface of antiferromagnetism and metallicity. This
idea, though attractive, is incorrect.  There is no SC
within the $\rho=1$ Hubbard Hamiltonian for any $U$ or anisotropy
\cite{Clay08a}.  Importantly, there exist many CTS in which the
superconducting transition is from an insulating state {\it different}
from antiferromagnetism, or where the triangular lattice of monomers
is not dimerized, which would also argue against any mechanism based
on the $\rho=1$ Hubbard Hamiltonian.  The only common features between
CTS superconductors are strong correlations, frustrated lattice of
molecules, and $\rho=0.5$.

The hole density in the superconducting hydrated cobaltate is more
elusive.  The original assumption that the Na-concentration determines
the hole density also in the hydrated material has been found to be
not true; some water molecules enter as H$_3$O$^+$, and the actual
$\rho$ in the superconductor is much smaller than the 0.65 that would
be guessed from the Na-concentration. There have been several reports
that SC occurs over a very narrow range of hole density, and that
maximum T$_c$ occurs at or very close to Co-ion valency 3.5+
\cite{Barnes05a}, corresponding to $\rho=0.5$.

{\it The common features between all organic superconducting CTS and
  the cobaltates then appear to be strong correlations, geometric
  lattice frustration and $\rho=\frac{1}{2}$.}  Yet another class of
compounds that share these features are inorganic spinels AB$_2$X$_4$.
The B sublattice forms a frustrated pyrochlore lattice and usually
consists of transition metal cations with partially filled t$_{2g}$
d-orbitals. Only four of the many spinel compounds are
superconducting, of which three have effective carrier density
$\rho=0.5$: LiTi$_2$O$_4$, CuRh$_2$S$_4$, and CuRh$_2$Se$_4$. In
LiTi$_2$O$_4$ there is one d-electron per two Ti$^{3.5+}$ ions.  The
Rh$^{3.5+}$ ions, like Co$^{3.5+}$, are also $\rho=0.5$.

Elsewhere we have shown that increased frustration within Eq.~1 leads
to an antiferromagnetism-to-spin singlet charge-ordered transition in
$\rho=0.5$ \cite{Dayal11a}, and that this same spin-singlet state can
undergo transition to a superconducting state with further increase in
frustration \cite{Mazumdar08a}. While further work on the proposed
superconducting transition is required, the attractive features of
this scenario are that the peculiar $\rho$-dependence of the above
diverse materials are explained within Eq.~1, and the possibility
exists that SC can be explained within a single plausible mechanism.


\begin{thebibliography}{00}



\bibitem{Powell06a} B. J. Powell and R. H. McKenzie, J. Phys. Condens. Matter {\bf 18}, R827 (2006) and
references therein.

\bibitem{Baskaran03a} G. Baskaran, Phys. Rev. Lett. {\bf 91}, 097003 (2003). 

\bibitem{Kumar03a} B. Kumar and B. S. Shastry,
Phys. Rev. B {\bf 68}, 104508 (2003). 

\bibitem{Motrunich04a} O. I. Motrunich and P. A. Lee, Phys. Rev. B {\bf 69}, 214516 (2004).

\bibitem{Wang03a} Y. Wang {\it et al.}, Nature {\bf 423}, 425 (2003).

\bibitem{ARPES} M. Z. Hasan {\it et al.}, Phys. Rev. Lett. {\bf 92}, 246402 (2004). H.-B. Yang
{\it et al.}, {\it ibid}, {\bf 95}, 146401 (2005). D. Qian {\it et al.}, {\it ibid}, {\bf 96},
216405 (2006).

\bibitem{Laverock07a} J. Laverock {\it et al.} have found $\rho$-dependent $e_g^\prime$-participation
from Compton scattering measurements, see Phys. Rev. B {\bf 76}, 052509 (2007).  

\bibitem{Torrance75a} J. B. Torrance, B. A. Scott, and F. B. Kaufman, Solid St. Commun. {\bf 17}, 1369 (1975).

\bibitem{Mazumdar86a} S. Mazumdar and S. N. Dixit, Phys. Rev. B {\bf 34}, 3683 (1986).

\bibitem{Huizinga79a} S. Huizinga {\it et al.} Phys. Rev. B {\bf 19}, 4723 (1979).

\bibitem{Comes79a} R. Comes and G. Shirane in {\it Highly Conducting  One-Dimensional Solids}, edited by
J. T. Devreese {\it et al.} (Plenum, New York, 1979), pp. 17-67.

\bibitem{Nad06a} F. Nad and P. Monceau, J. Phys. Soc. Jpn. {\bf 75}, 051005 (2006) and references therein.
D. S. Chow {\it et al.}, Phys. Rev. Lett. {\bf 85}, 1698 (2000). 

%

\bibitem{Foo04a} M. L. Foo {\it et al.}, Phys. Rev. Lett. {\bf 92}, 247001 (2004).

\bibitem{RecentCoO2a} G. Lang {\it et al.}, Phys. Rev. B {\bf 78}, 155116 (2008). T. Motohashi {\it et al.}, {ibid},
{\bf 83}, 195128 (2011). V. Brouet {\it et al.}, {ibid}, {\bf 76}, 100403(R) (2007).

\bibitem{Mazumdar83a} S. Mazumdar and A. N. Bloch, Phys. Rev. Lett. {\bf 50}, 207 (1983).

\bibitem{Monceau01a} P. Monceau, F. Y. Nad and S. Brazovskii Phys. Rev. Lett. {\bf 86}, 4080 (2001).
H. Seo {\it et al.}, J. Phys. Soc. Jpn. {\bf 75}, 051009 (2006). R. T. Clay, R. P. Hardikar and S. Mazumdar, Phys. Rev. B
{\bf 76}, 205118 (2007).

%
\bibitem{Li11a} H. Li, R. T. Clay and S. Mazumdar, Phys. Rev. Lett. {\bf 106}, 216401 (2011).

\bibitem{Choy07a} T. P. Choy, D. Galanakis, and P. Phillips, Phys. Rev. B {\bf 75}, 073103
(2007) and references therein.

\bibitem{Barnes05a} P. W. Barnes {\it et al.} Phys. Rev. B {\bf 72}, 134515 (2005). H. Sakurai {\it et al.},
{\it ibid}, {\bf 74}, 092502 (2006). M. \protect{Ba\~nobre-L\'opez} {\it et al.}, J. Am. Chem. Soc. {\bf 131}, 9632 (2009).
T. Shimojima {\it et al.}, Phys. Rev. Lett. {\bf 97}, 267003 (2006).

\bibitem{Clay08a} R. T. Clay, H. Li and S. Mazumdar, Phys. Rev. Lett. {\bf 101}, 166403 (2008).

\bibitem{Dayal11a} S. Dayal {\it et al.} Phys. Rev. B {\bf 83}, 245106 (2011).

\bibitem{Mazumdar08a} S. Mazumdar and R. T. Clay, Phys. Rev. B {\bf 77}, 180515 (2008).

\end{thebibliography}
\end{document}